\RequirePackage{ifpdf}
\documentclass{PoS}

\usepackage[authoryear,square]{natbib}
\bibpunct{(}{)}{;}{a}{}{,}

\title{Imaging HII Regions from Galaxies and Quasars During Reionisation with SKA}

\ShortTitle{Imaging Reionisation with SKA}

\author{
\speaker{J. Stuart B. Wyithe}, 
Paul M. Geil,
Hansik Kim
\\ 
School of Physics, The University of Melbourne, Parkville, Vic 3010, Australia
\\
E-mail: \email{swyithe@unimelb.edu.au}
}

\abstract{
The ionisation structure of the Intergalactic Medium (IGM) during reionisation is sensitive to the unknown galaxy formation physics that prevailed at that time. This structure introduces non-Gaussian statistics into the redshifted 21~cm fluctuation amplitudes that can only be studied through tomographic imaging, which will clearly discriminate between different galaxy formation scenarios. Imaging the ionisation structure and cosmological HII regions during reionisation is therefore a key goal for the SKA. For example, the SKA1-LOW baseline design with a 1~km diameter core will resolve HII regions expected from galaxy formation models which include strong feedback on low-mass galaxy formation. Imaging the smaller HII regions that result from galaxy formation in the absence of SNe feedback will also be possible for SKA1-LOW in the later stages of reionisation, but may require the greater sensitivity of SKA early in the reionisation era. In addition to having baselines long enough to resolve the HII regions, the field of view for SKA1-LOW reionisation experiments should be at least several degrees in order to image the largest HI structures towards the end of reionisation. The baseline design with 35 meter diameter stations has a  field of view within a single primary pointing which is sufficient for this purpose.
}

\FullConference{
Advancing Astrophysics with the Square Kilometre Array\\
June 8-13, 2014\\
Giardini Naxos, Italy}

\newcommand{\skipthis}[1]{}

\begin{document}

\section{Introduction}

Redshifted 21~cm emission from HI in the intergalactic medium (IGM) offers several probes of the reionisation epoch. Owing to the low signal-to-noise expected for first generation telescopes, including the MWA and LOFAR, most attention has been focussed on statistical observations of the power-spectrum of 21~cm fluctuations together with its evolution. However several other observational signatures become more powerful probes of reionisation using the imaging capability of SKA1-LOW. These include $i)$ the cross-correlation of 21~cm emission with galaxies, which would directly probe the connection between reionisation and the sources of ionising radiation \citep{WL07,furlanetto2007,lidz2009,Park2013},  $ii)$ the probability distribution of intensity fluctuations, which contains additional information about the non-Gaussian intensity fluctuations during reionisation~\citep{WL07,Harker2009a,Barkana2008}, $iii)$ the observation of individual HII regions, which will probe galaxy formation physics, and $iv)$ the observation of quasar-dominated HII regions, which will probe quasar emission geometries  as well as the evolution of the neutral IGM \citep{wl2004b,kohler2005,valdes2006}.

While the cross-correlation of 21~cm emission with galaxies may be possible with first generation arrays using the statistics of the cross power-spectrum \citep{WL07,furlanetto2007,lidz2009}, direct observation of the correspondence between galaxies and ionisation structure will provide unambiguous evidence of the role of galaxies in reionisation. An important prediction from the various modelling studies concerns whether over-dense or under-dense regions become ionised first. Standard calculations of the expected cross-correlation between the distribution of galaxies and the intergalactic 21~cm emission at high redshifts show that over-dense regions will be ionised early \citep[e.g.][]{WL07}, leading to an anti-correlation between 21~cm emission and the galaxy population. Redshifted 21~cm surveys should be able to discriminate between "outside-in" and "inside-out" reionisation, even when combined with galaxy surveys of only several square degrees.

The distribution of fluctuation amplitudes is expected to be Gaussian during the early phases of reionisation. At these early times the power-spectrum represents the natural quantity to describe fluctuations in 21~cm emission, since it contains all the statistical information. However the distribution of fluctuation amplitudes becomes non-Gaussian as reionisation progresses~\citep{M2006,WM07,harker2009}. This is particularly the case on small scales once HII regions have formed~\citep{furl2004b}. As a result, the power-spectrum does not provide a complete statistical description of the reionisation process, and instead the full probability distribution of intensity fluctuations should be considered.  For example, additional information could be captured by measuring skewness in the probability distribution~\citep{WM07,harker2009}. However, unlike the power-spectrum, the probability distribution for intensity fluctuations must be derived from images of the ionisation structure.

In this chapter we focus on the prospects for studying galaxy formation through the imaging of HII regions, both in the typical IGM as well as around quasars using SKA1-LOW. The direct imaging of the ionised structure during reionisation would be the most unambiguous measurement of reionisation. This represents the "holy grail" of 21~cm reionisation experiments, and is a key goal for the SKA. The importance of imaging with SKA  for capturing the physics of reionisation is discussed in more detail in \citet{Mellema}.  Discussion of the science potential of SKA1-LOW for imaging the ionised structure during reionisation requires modelling of ionised regions on very large scales, and we therefore begin this chapter with a brief discussion of the semi-numerical method used to calculate ionisation structure. A more detailed discussion of issues associated with simulation of reionisation can be found in \citet{Illiev}.

\section{Semi-numerical model for reionisation}

In recent years approximate, but efficient semi-numerical methods for simulating the reionisation process have been developed \citep{MF07} in order to overcome the limitations of both analytic and numerical methods. These extend prior work \citep{bond1996a,zahn2007} to estimate the ionisation field based on a catalogue of sources by applying a filtering technique based on an analytic HII region model \citep{furl2004b}. The resulting structure of the ionisation field is similar for different variations on the filtering method as well as for full radiative transfer, implying that semi-numerical models can be used to explore a larger range of reionisation scenarios than is possible with current numerical simulations \citep{MF07}.  Alternatively, rather than adopt a filtering scheme, ionising sources can be used used to generate HII regions based on the spherically averaged radial density profile~\citep{thomas2009}. A feature of this model is that overlap of neighbouring HII regions is treated self-consistently with respect to photon conservation. The model can also be easily adapted to compute ionisation structure for more complex galaxy formation models.

We use a combination of these techniques implemented within the semi-analytic model GALFORM \citep[e.g.][]{Lagos12} in order to study the effect of galaxy formation properties on the ionisation structure of the IGM \citep{Kim12}. Beginning with a relatively small ($\sim100$ Mpc$/h$ cubed) volume simulation of the ionisation structure in which the physics of the lowest mass sources thought to dominate reionisation are resolved, this method can be extended to very large volumes using the statistical distribution of ionising flux as a function of large-scale ($\sim$Mpc) over density (Kim 2014, in prep). This distribution is used to populate a much larger low resolution volume via a Monte-Carlo technique, thus producing simulations of ionisation structure in cubic Gpc volumes that contain the SKA1-LOW field of view, while retaining the effects of physics from the smallest-scale galaxies.

\subsection{HII regions and galaxy formation}

\begin{figure*}
\begin{center}
\includegraphics[width= 15.5cm]{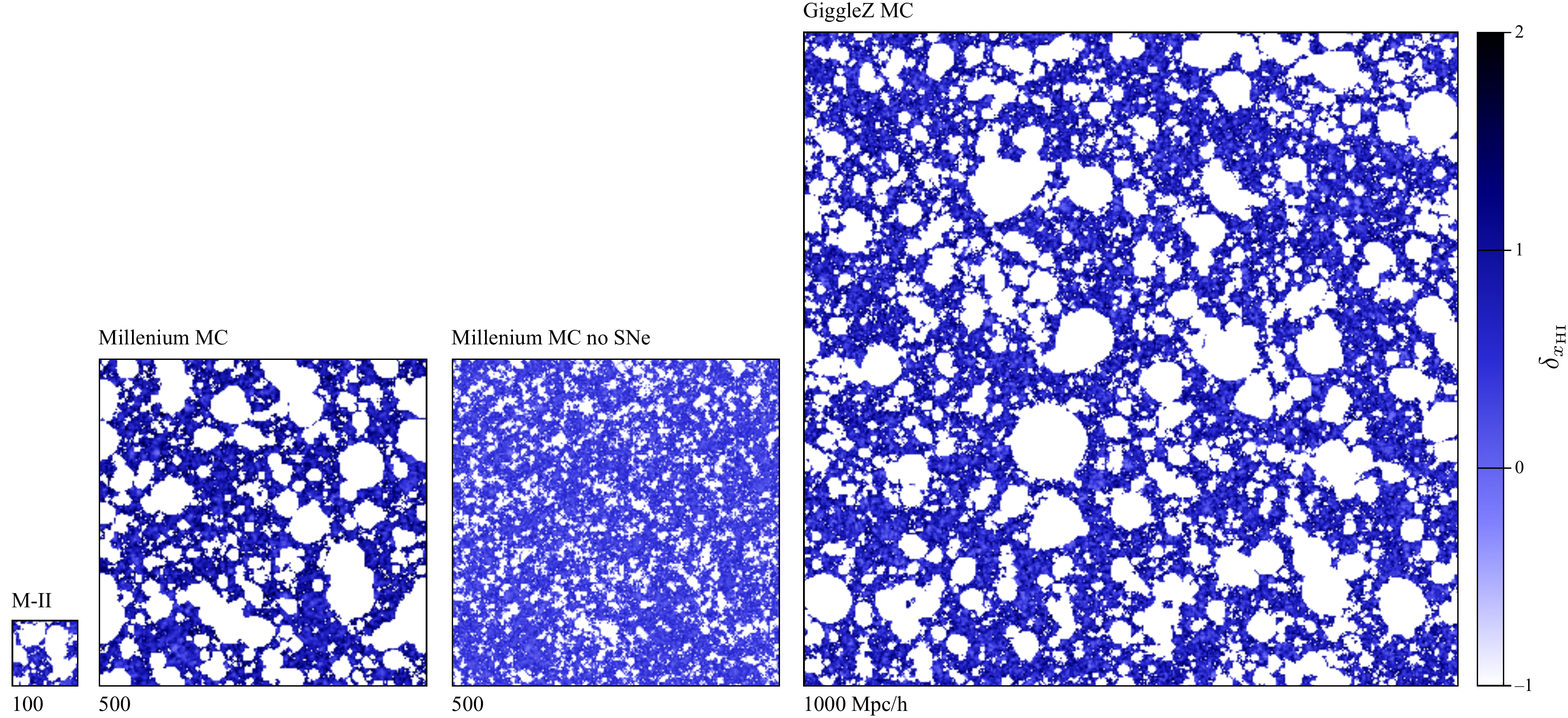} 
\caption{Simulations of the HII region structure in the Millennium-II simulation (left), with Monte-Carlo (MC) extension to the Millennium simulation (second and third panels) and the Gigglez 1Gpc cubed simulation (right panel). In each case simulations are shown for the default GALFORM galaxy formation model with SNe feedback, and reionisation was assumed to have progressed to a neutral fraction of 0.45. In the case of the Millennium simulation we also show the scenario where the galaxy formation model does not include SNe feedback (third panel from left).  The model slices have a depth 2Mpc$/h$.}
\label{HankHII}
\end{center}
\end{figure*}

Examples of slices through simulated volumes of ionisation structure are shown in Figure~\ref{HankHII}. Several different volumes are shown at $z=7.272$ to illustrate the scales involved. All simulations have the same mass-averaged neutral hydrogen fraction of $\langle x_{\rm HI}\rangle\sim0.45$. From left to right we show the HII region structure resulting from the default GALFORM galaxy formation model implemented within the 100 Mpc$/h$ cubed volume of the Millennium-II simulation \citep{Kim12}, as well as the extension via our Monte-Carlo method (Kim 2014 in prep) to the 500 Mpc$/h$ cubed volume of the Millennium simulation and 1Gpc$/h$ cubed volume of the GiggleZ simulation~\citep{GiggleZ}. Note that ionised volumes that are nearly as large as the 100 Mpc$/h$ cubed Millennium-II simulation can be seen in the larger 500 Mpc$/h$ cubed Millennium and 1 Gpc$/h$ cubed GiggleZ simulation volumes. These very large HII regions indicate that Gpc-scale volumes are required to fully understand the large-scale power of reionisation~\citep{Iliev13}.

In addition to the default GALFORM galaxy formation model in which SNe feedback plays a significant role in regulating star formation, for the Millennium 500 Mpc$/h$ cubed simulation we also show a second example in which SNe feedback is absent. In each case the galaxy formation model produces an acceptable description of the high redshift galaxy luminosity function, and the mean neutral fraction of the IGM in both models is forced to be equal by construction. In the case of a galaxy formation model without SNe feedback, we find that much smaller HII regions are produced owing to the smaller luminosity-weighted mass of the ionising galaxies~\citep{Kim12}, requiring imaging at greater resolution in order to study the structure of the HII regions.

\subsection{Quasar HII regions}
	 	 
\begin{figure}
\begin{center}
\includegraphics[width = 14.5 cm]{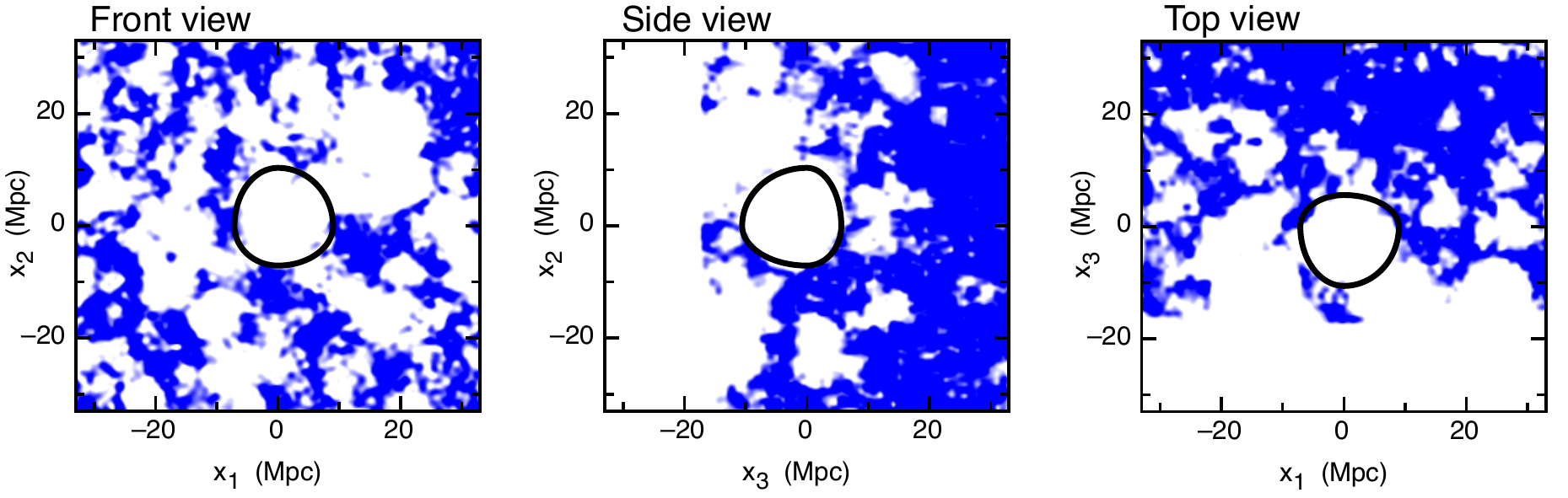}
\caption{A quasar HII region in an evolving IGM \citep[figure adapted from ][]{GWPO08}. Three aspects of the HII region are shown, with slices through the centre of the box when viewed from the front, top and side. The quasar was assumed to contribute ionisation equivalent to an HII region of radius $R_{\rm q}=34$ co-moving Mpc, and to be centred on $z=6.65$, which is also the redshift at the centre of the simulation box. Each slice is $6\,$Mpc thick, which corresponds to $\sim3\,$MHz along the $x_3$-axis (units in this figure are shown in physical Mpc). In observed units, the cube is $\sim3.3$ degrees on a side and $33\,$MHz deep. The shape of the HII region is also plotted~\citep[see][for details]{GWPO08}. The mass-averaged IGM neutral fraction was assumed to be $0.15$ at the quasar redshift. }
\label{Wyithe1}
\end{center}
\end{figure}

Spectra of several of the most distant known quasars at $z\sim6$ exhibit evidence for the presence of an HII region in a partially neutral IGM \citep[e.g.][]{cen2000,wl2004b}, although this interpretation remains uncertain~\citep{lidz2006, bolton2007}. Recently, stronger evidence for a neutral IGM has been found at $z\sim7$~\citep{Bolton2011}, where spectra show the possible detection of an IGM-generated Ly$\alpha$ damping wing in addition to a Ly$\alpha$ absorption trough.  The redshifted 21~cm observation of quasar HII regions will probe quasar physics as well as the evolution of the neutral gas \citep{wl2004b,kohler2005}. For example, Figure~\ref{Wyithe1} shows a semi-numerical model of the evolving 3-dimensional ionisation structure of the IGM within a 500 Mpc cubed volume around a luminous $z>6$ quasar~\citep{GWPO08} observed near the end of reionisation.  The evolution of the IGM, which is assumed to be quite rapid, is clearly seen in this figure with the percolation process completing between the "back" of the box and the "front" of the box. We note that in difference to galaxy HII regions which are driven by many sources over a long period of time, during the early phase quasar driven HII regions expand with a relativistic speed \citep{WL04}. Consequently, their measured sizes along and transverse to the line-of-sight should have different observed values due to relativistic time delay. A combined measurement of these sizes could therefore be used to directly constrain the neutral fraction of the surrounding IGM as well as the quasar lifetime \citep{WL04}. The figure also illustrates asymmetries perpendicular to the line of sight, which are a result of expansion of the spherical quasar driven HII region into an in homogeneously ionised IGM.

\section{Imaging sensitivity of SKA1-LOW}

\label{sensitivity}
In this section we estimate the sensitivity of the proposed SKA1-LOW baseline design with respect to imaging of cosmological HII regions. The baseline design for the SKA1-LOW will consist of 949 stations, of which approximately 400 are located within a uniform station density profile within a 500~m radius, with the remaining stations distributed in equal area per logarithm of radius in spiral arms beyond the core. Each station consists of 256 dual-polarisation antenna elements within a diameter of 35 m. The system temperature at $\nu<200$\,MHz is due to the combined sky and receiver temperatures, with a value  $T_{\rm sys}= T_{\rm sky} + T_{\rm rcvr}$, where  $T_{\rm sky}\sim60\left[\lambda/m\right]^{2.55}$\,K and $T_{\rm rcvr} = 0.1T_{\rm sky} + 40~K$. Simulations described in the document {\em SKA1 Imaging Science Performance}\footnote{Braun,
R., "SKA1 Imaging Science Performance", Document no. SKA-TEL-SKO-DD-XXX Revision A Draft 2} yield a noise level that should be within a factor of 2 of the natural instrument sensitivity. The angular resolution corresponding to the 500~m radius core at 171 MHz is $\theta_{\rm b} = 7.3$ arc minutes, and the primary beam width for the 35 m station is $\Omega = 3.5$ degrees (defined as centre to first null). For reference, the resulting rms noise in an image constructed in the manner described, for a frequency channel $\Delta\nu$ has the form
\begin{equation}
\label{Tb}
\Delta T_{\rm b} \approx 0.05\,{\rm mK} + 0.66\,{\rm mK}\left(\frac{1+z}{8.5}\right)^{2.55}\left(\frac{\Delta\nu}{1\,\mathrm{MHz}}\frac{t_{\mathrm{int}}}{1000\,\mathrm{hr}}\right)^{-1/2}\left(\frac{\theta_{\rm b}}{7'}\right)^{-2}.
\end{equation}

Radio interferometers do not directly image the full range of spatial scales, but rather measure a frequency-dependent, complex visibility for each frequency channel and baseline $\mathbf{U}$ in their configuration. The measured visibility is a linear combination of signal and noise, with the latter proportional to the square-root of the effective fraction of the array that can observe a particular visibility $\mathbf{U}$, which is in turn proportional to the number density of baselines $n(\mathbf{U})$ that can observe the visibility~\citep{mcquinn2006}. We simulate the thermal noise in a 3-dimensional visibility-frequency cube~\citep{GWPO08}, and then perform a 2-dimensional inverse Fourier transform in the $uv$-plane for each binned frequency in the bandwidth, which gives a realisation of the system noise in the image cube (i.e. sky coordinates). We scale the noise in each channel to have the variance described by equation~(\ref{Tb}). We construct the image using all baselines of up to 1000~m, including both core stations and stations along the spiral arms.

\section{Simulated images of ionised structure with SKA1-LOW}

\begin{figure*}
\begin{center}
\includegraphics[width= 7.4cm]{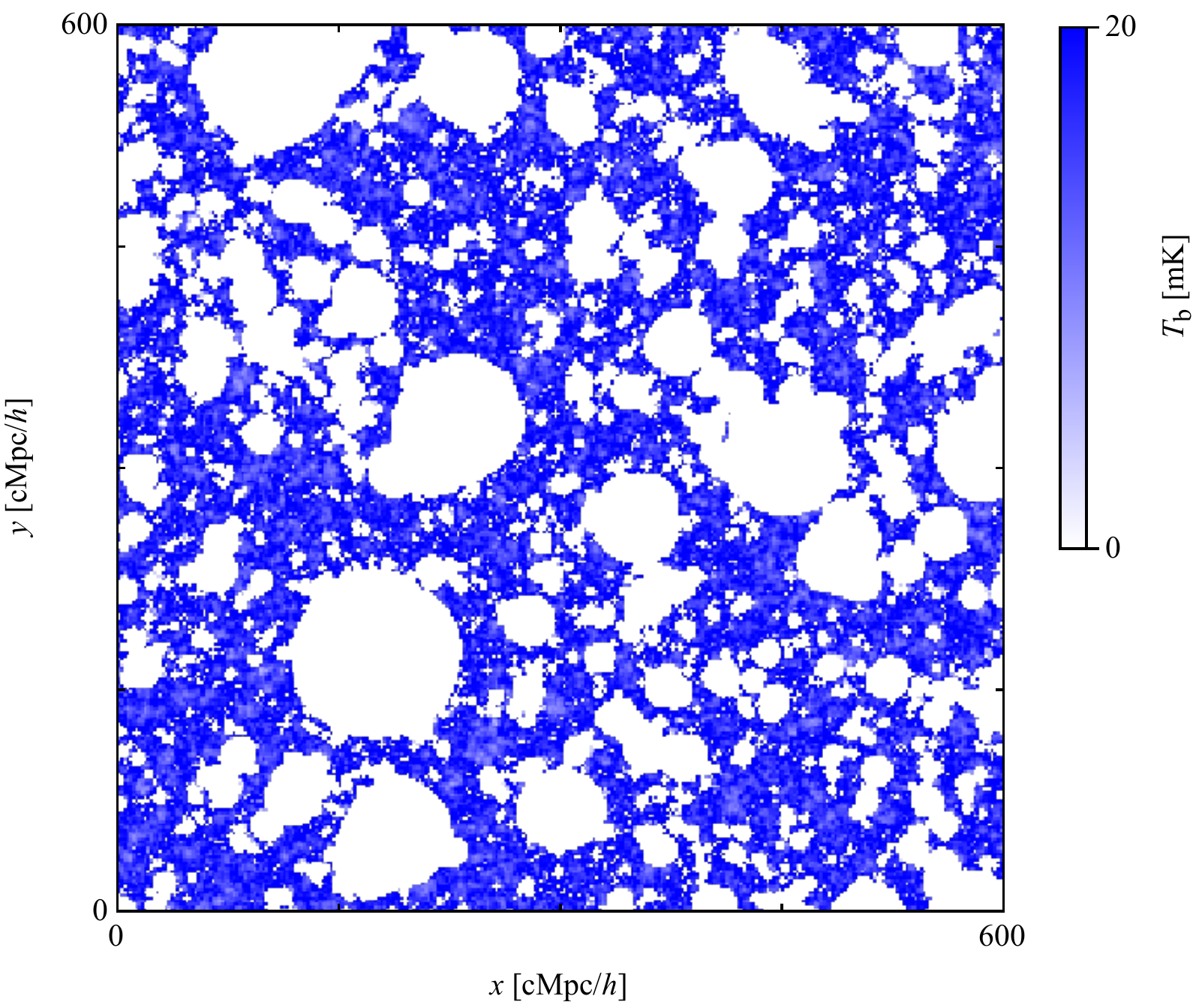}  
\includegraphics[width= 7.4cm]{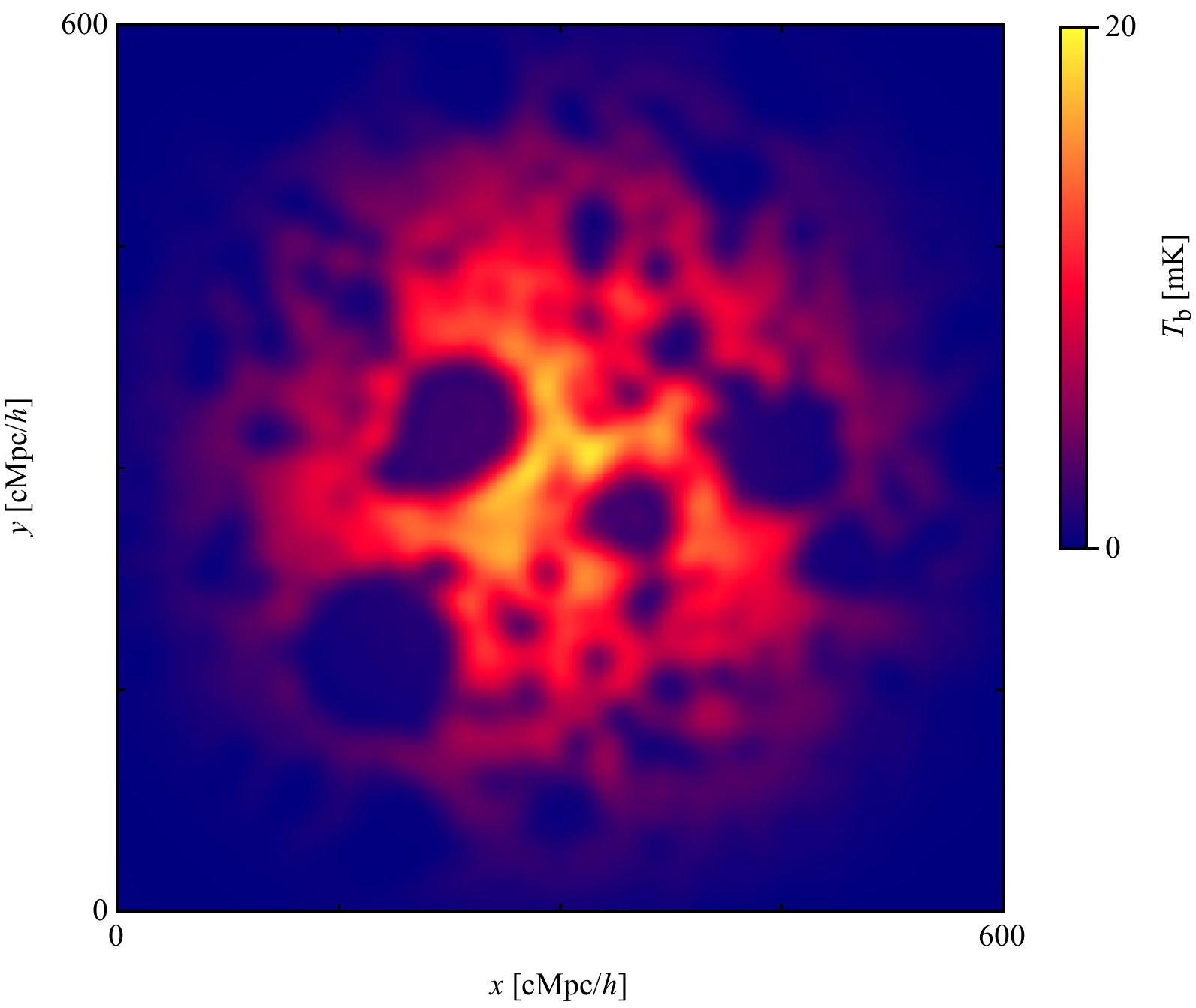} 
\includegraphics[width= 7.4cm]{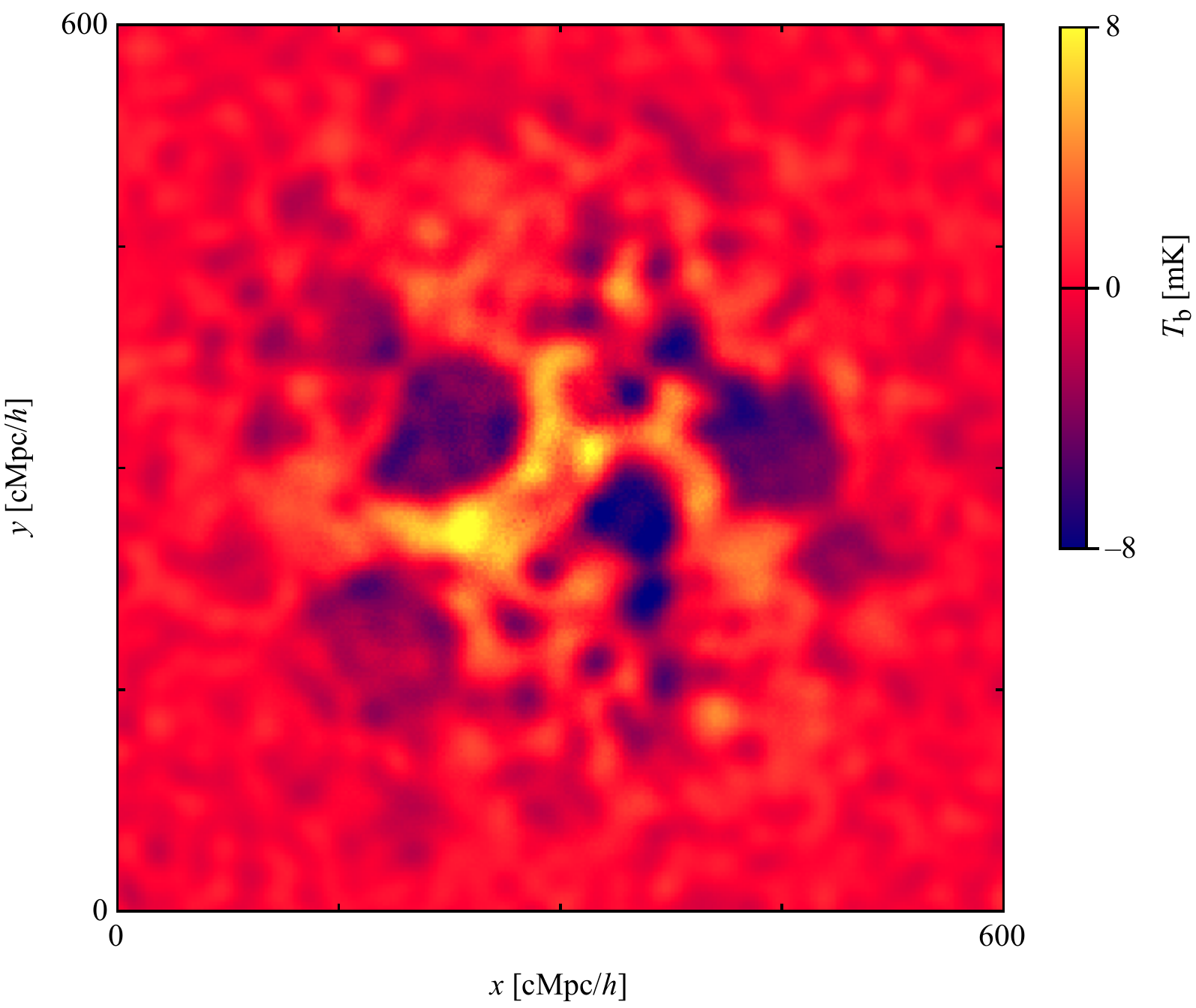}
\includegraphics[width= 7.4cm]{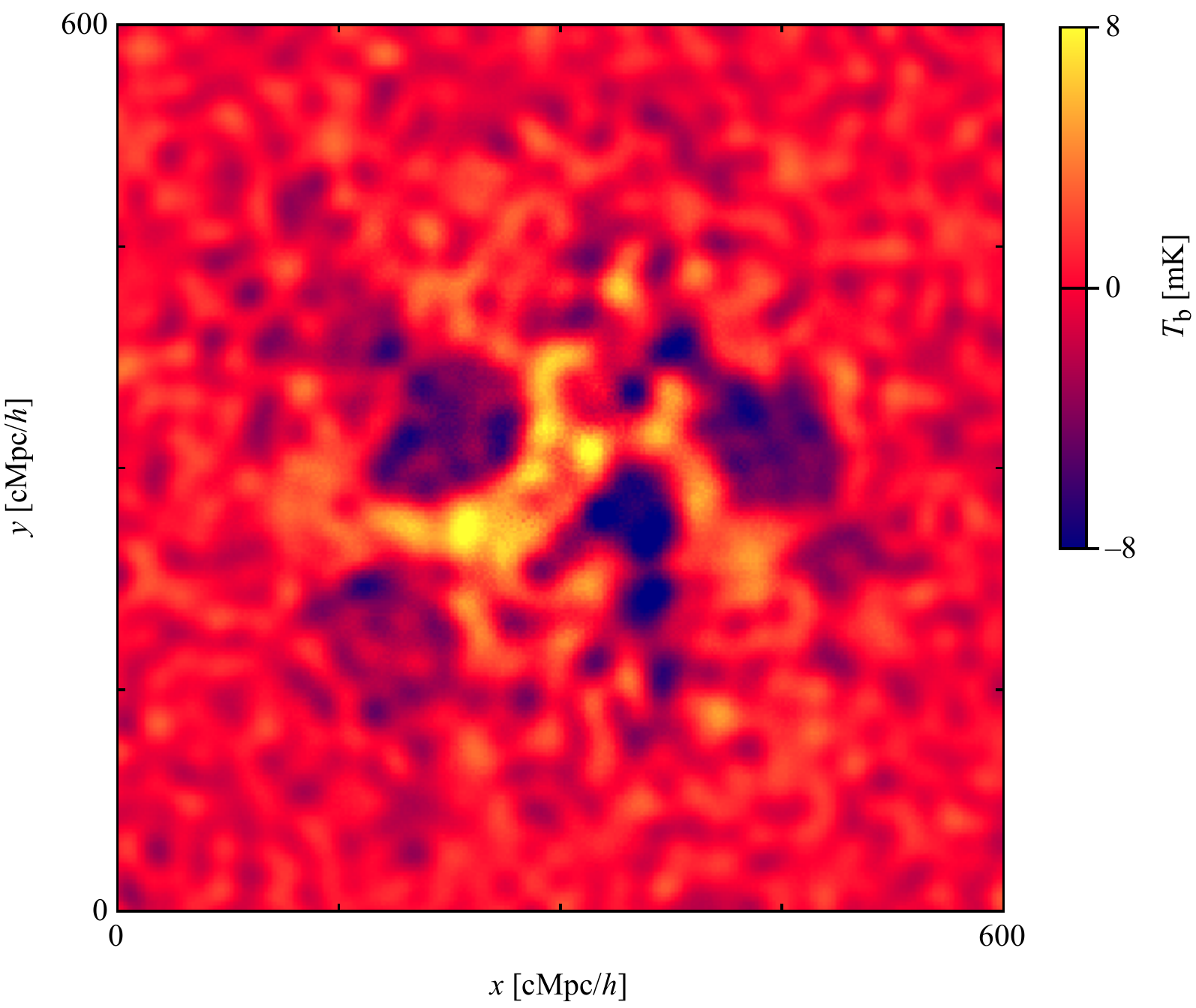}
\caption{Simulations of the SKA1-LOW response to the ionisation structure in the Gigglez 1~Gpc$/h$ simulation in which galaxy formation is assumed to include an efficient SNe feedback, and reionisation has progressed to a neutral fraction of 0.45. The upper-left panel shows the model slice of depth 2~Mpc$/h$ (corresponding to $171$~kHz at $z=7.27$). The upper-right panel shows an image of this slice assuming an estimate of the primary beam gain, accounting only for the power on scales that can be observed by baselines shorter than 1000~m (corresponding to the diameter of the core). The lower-left and lower-right panels show simulations of observed maps (without a primary beam correction) assuming the base-line SKA1-LOW with 1000~hr integration and an early deployment where the collecting area is decreased by a factor of 2. The model observations have a depth of 1.2~MHz, which corresponds to 14~Mpc$/h$  along the line-of-sight, at a central frequency of 173 MHz for HI at $z=7.27$, and is equivalent to the FWHP of the synthesised beam.   } 
\label{HankHII_obs}
\end{center}
\end{figure*}

In Figure~\ref{HankHII_obs} we estimate the SKA1-LOW response to the ionisation structure in the GiggleZ 1Gpc$/h$ cubed simulation in which galaxy formation is assumed to include an efficient SNe feedback. The top-left panel shows the model slice of depth 2 Mpc$/h$, which corresponds to 171 kHz along the line-of-sight, at a central frequency of 175MHz for HI at $z=7.27$. As in Figure~\ref{HankHII}, our model has a neutral fraction of 0.45 at this redshift. In addition to the properties of the image noise, the array configuration directly impacts the features of the image that can be measured. On scales where the density of baselines is low or zero, features in the image cannot be observed. This includes image power on both small and large scales. Finite visibility coverage, determined by the baseline distribution, therefore truncates the visibilities that make up the observed image. Thus, in addition to dictating the behaviour of noise, the array configuration determines which properties of the HII regions can be imaged. The upper-right panel of Figure~\ref{HankHII_obs} shows an image of the simulation slice including only power on those scales measured by baselines of up to 1000~m. Here we have assumed an estimate of the primary beam gain using the Fourier transform of a filled circle aperture.

\begin{figure*}
\begin{center}
\includegraphics[width= 7.4cm]{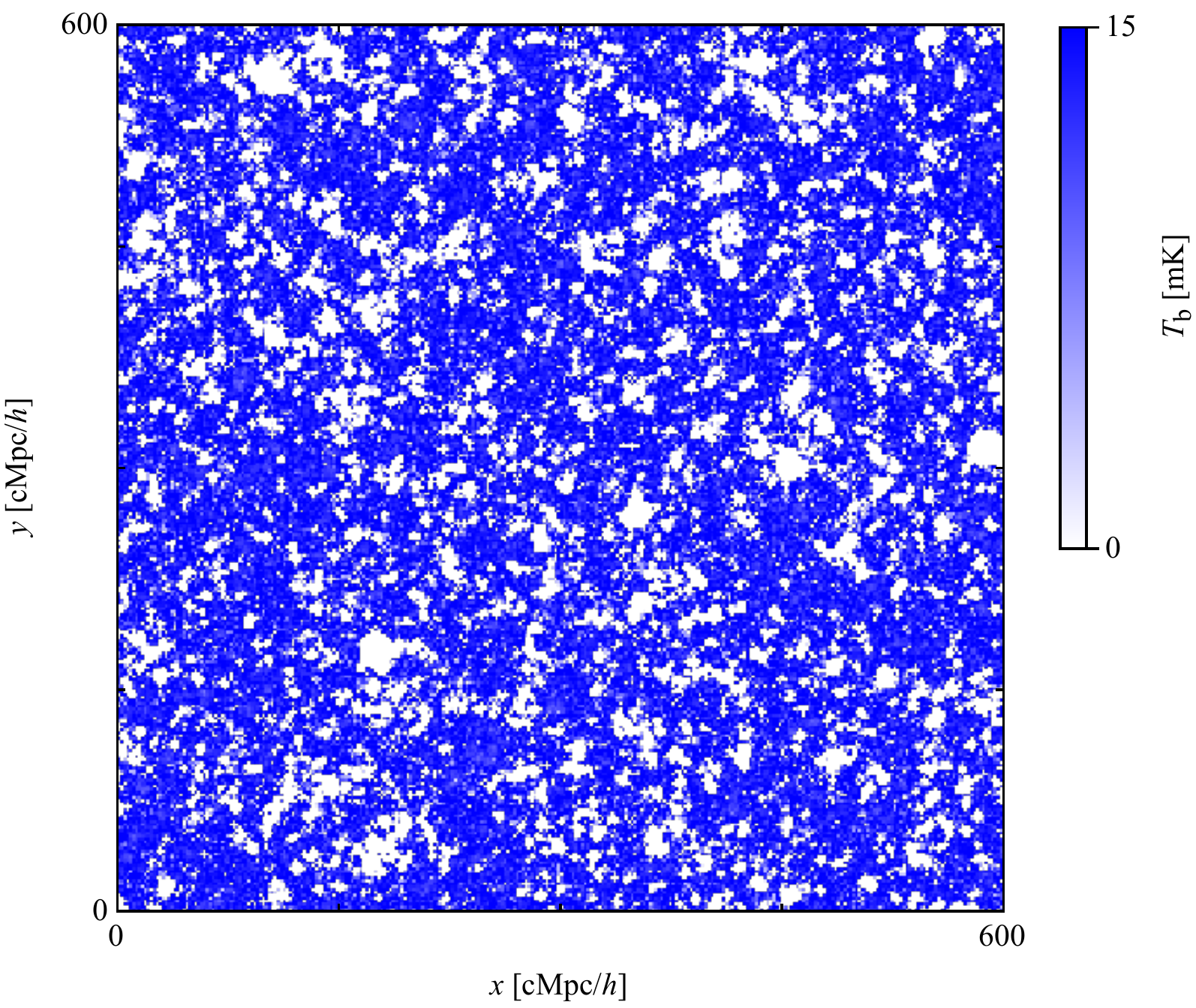}  
\includegraphics[width= 7.4cm]{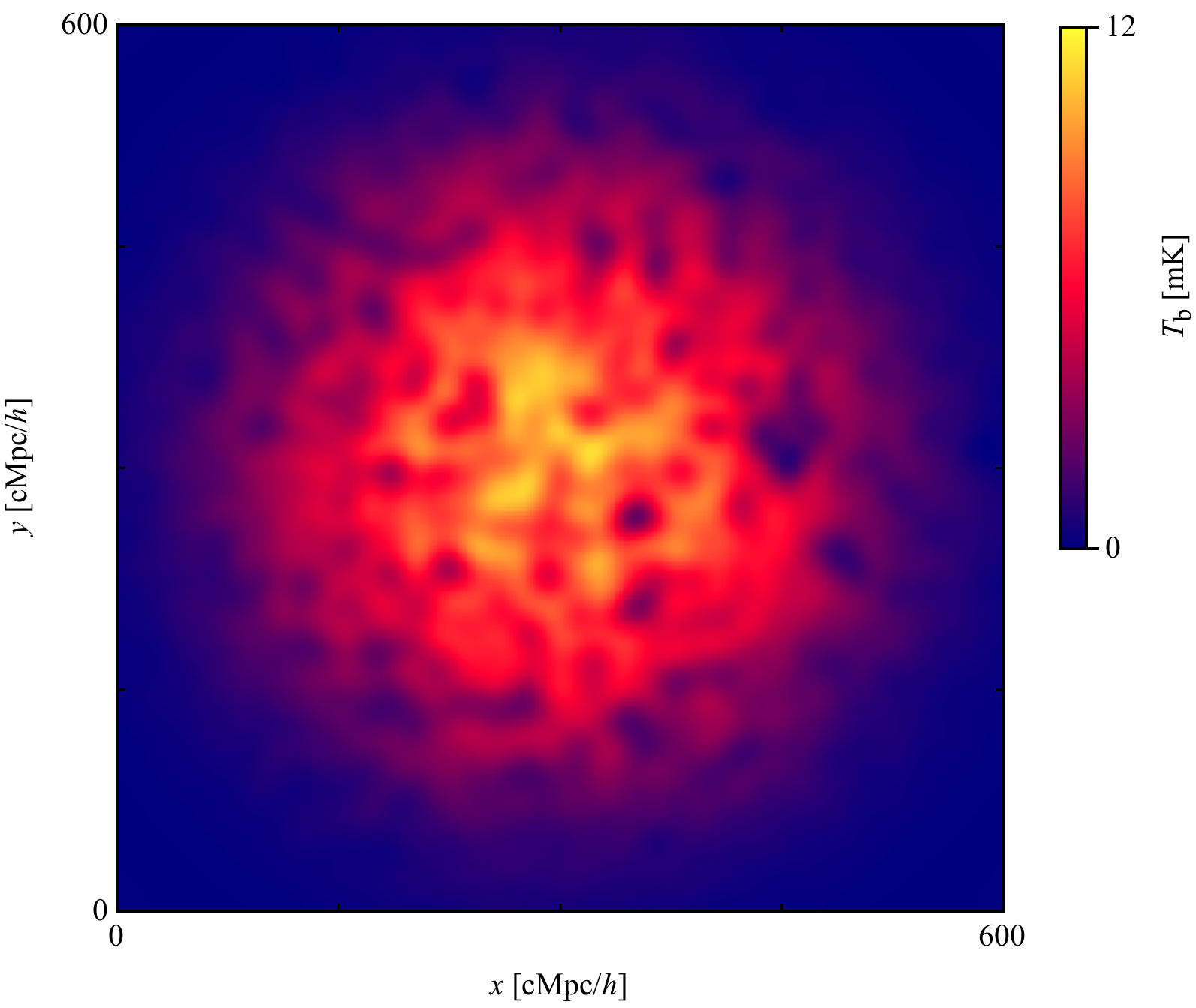} 
\includegraphics[width= 7.4cm]{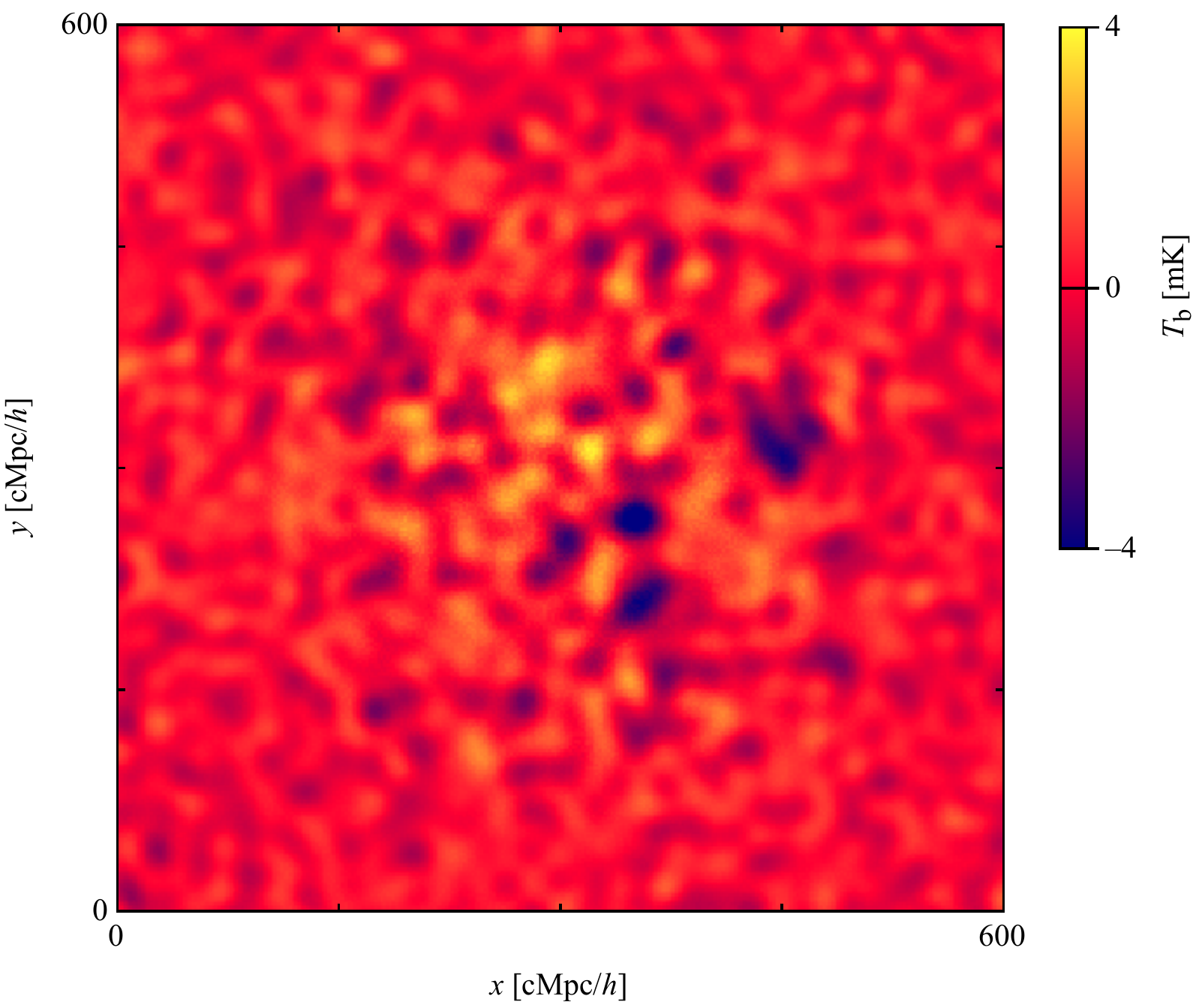}
\includegraphics[width= 7.4cm]{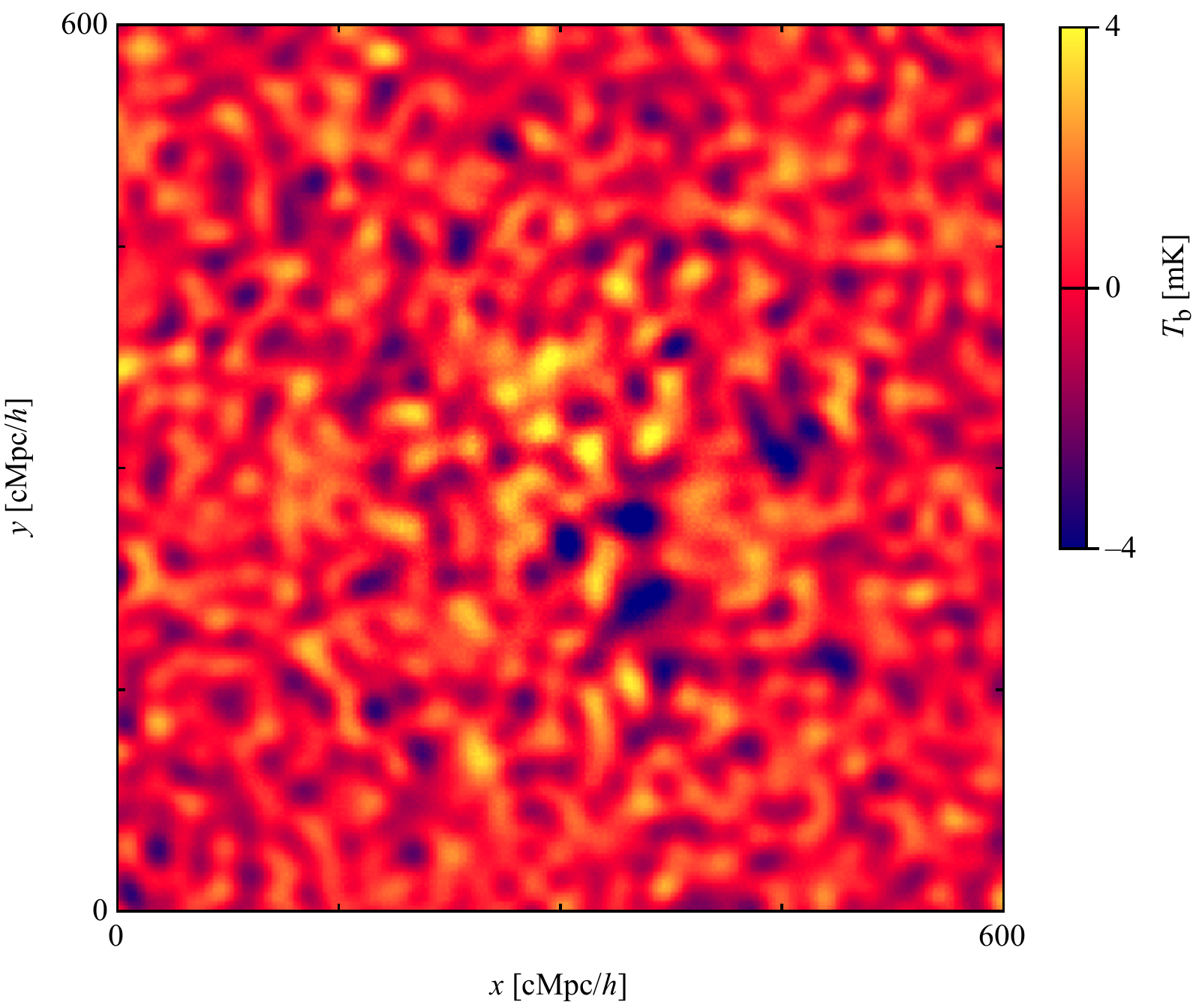}
\caption{Simulations of the SKA1-LOW response to the ionisation structure in the GiggleZ 1Gpc$/h$ simulation in which star formation is assumed to proceed in the absence of a strong SNe feedback. Details are as per Figure~3. }
\label{HankHII_obs_NOSN}
\end{center}
\end{figure*}

\begin{figure*}
\begin{center}
\includegraphics[width= 7.1cm]{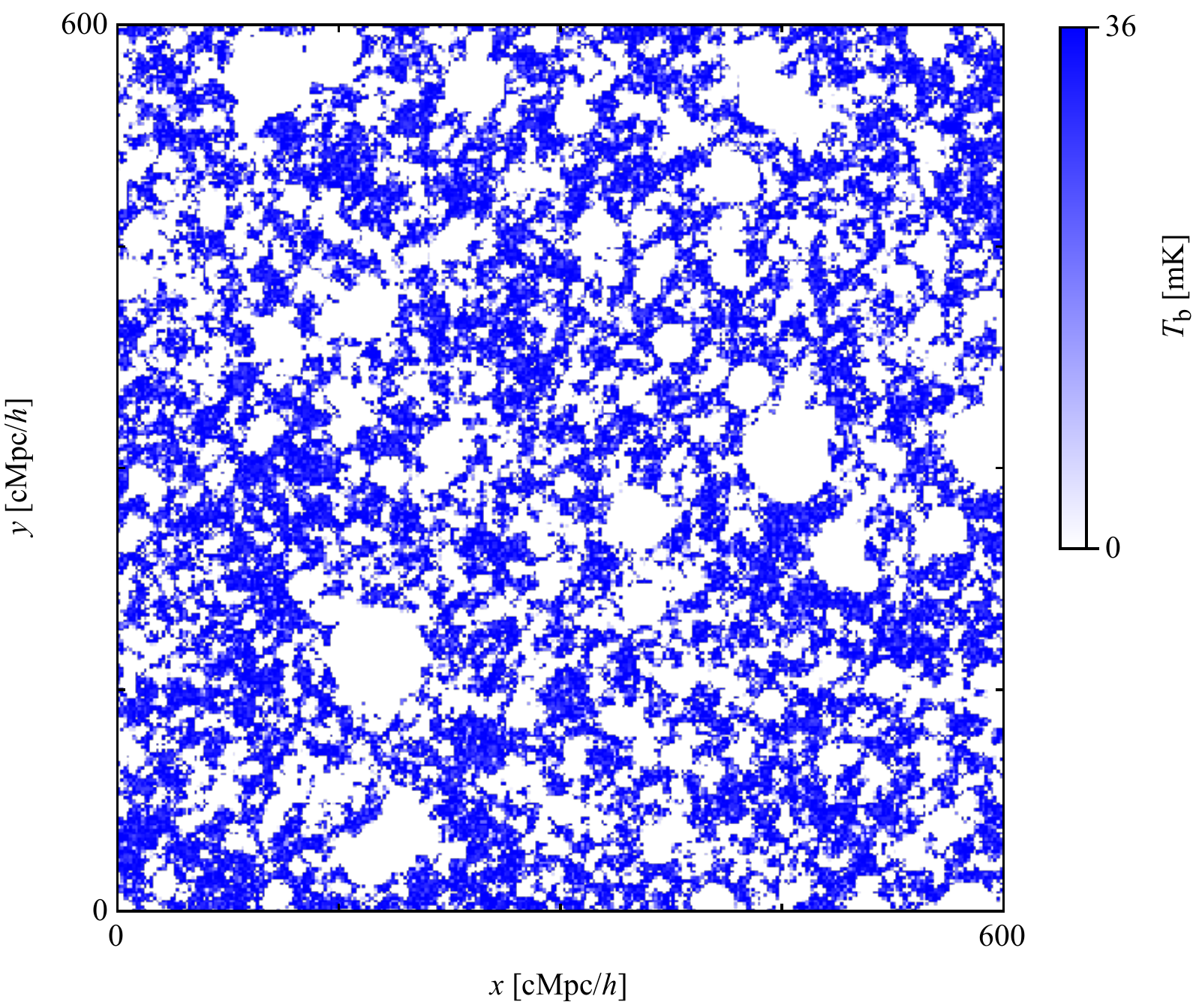}
\includegraphics[width= 7.1cm]{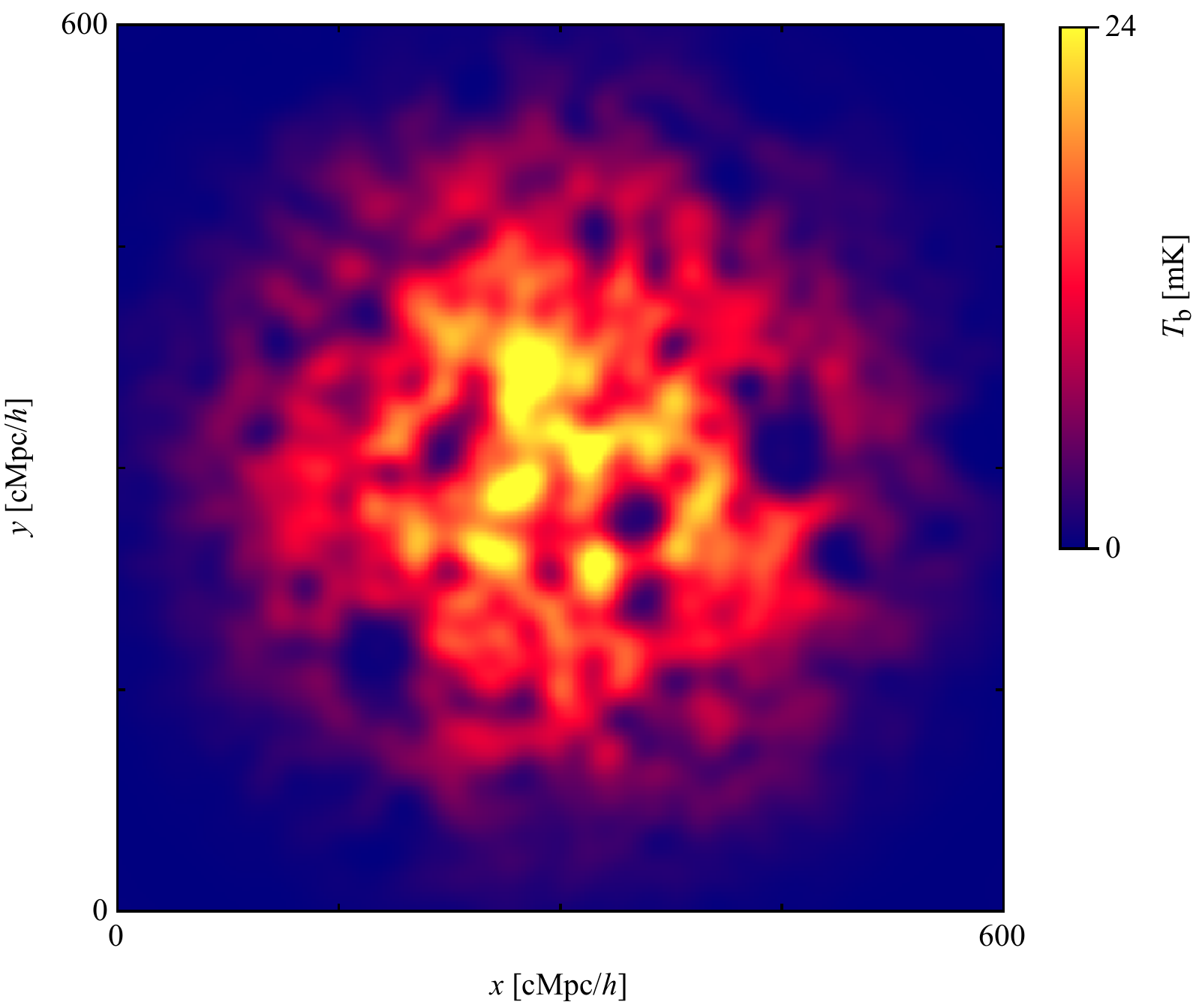}
\includegraphics[width= 7.1cm]{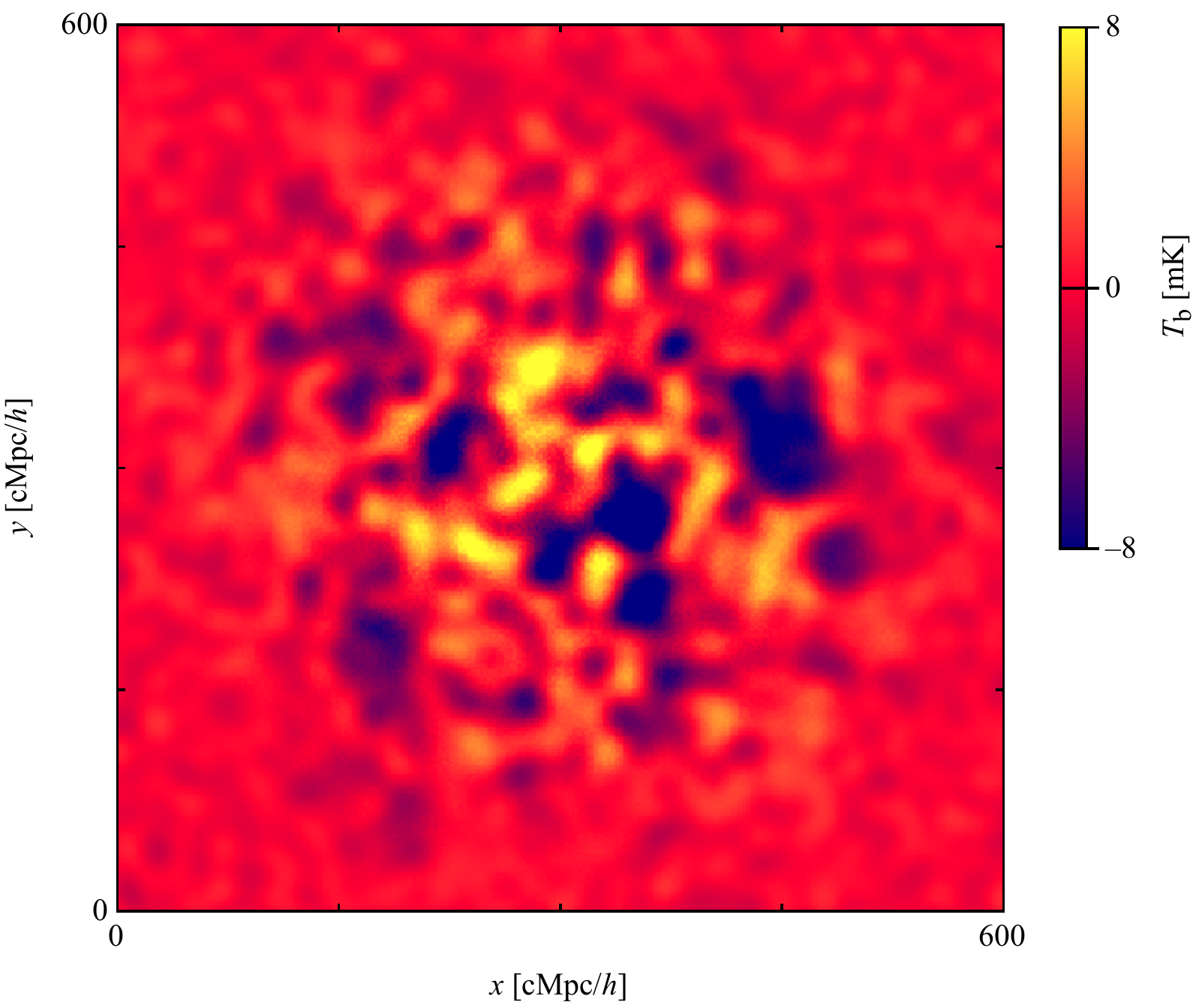}
\includegraphics[width= 7.1cm]{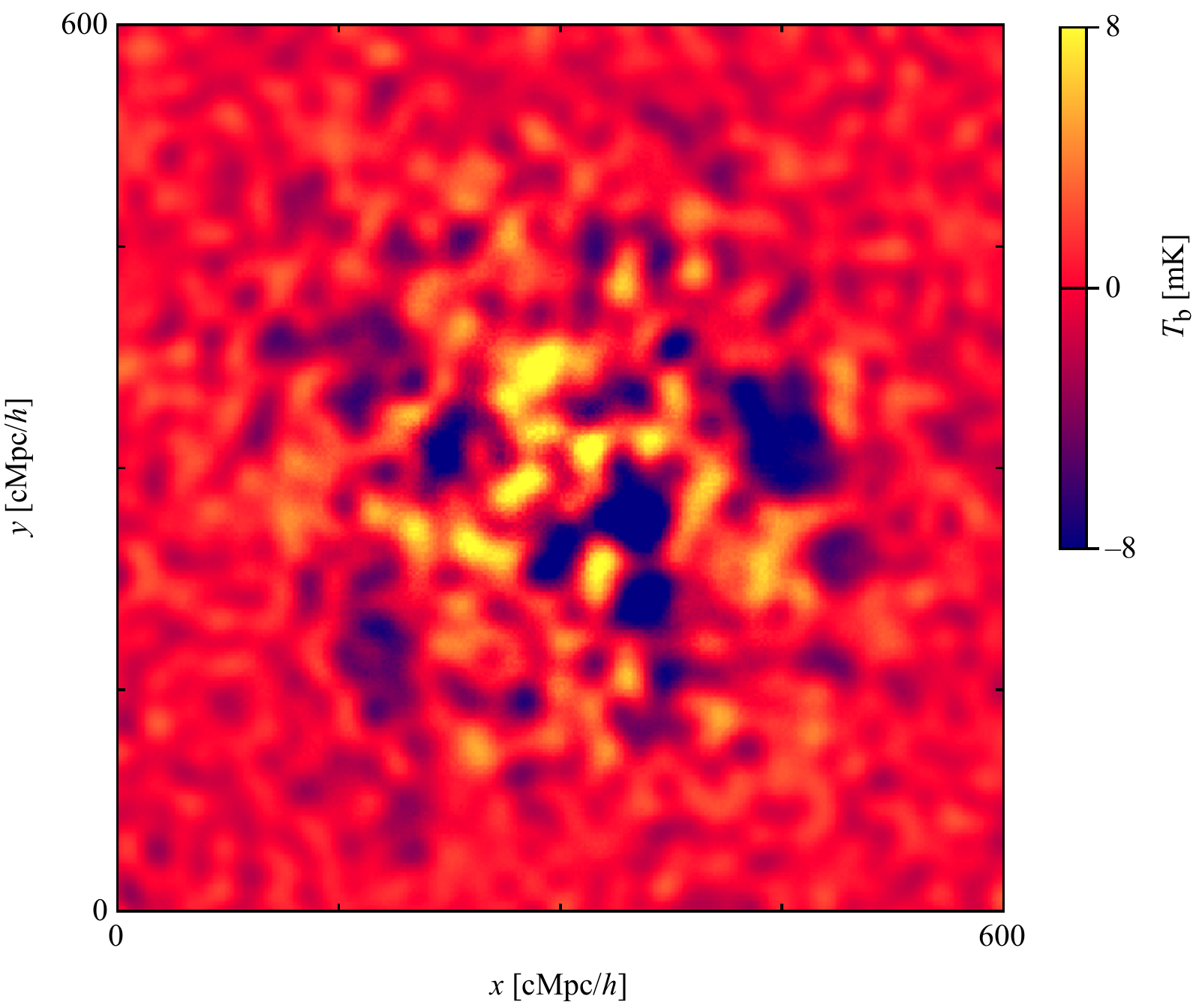}
\caption{As per Figure~4, but assuming a more advanced stage of reionisation with a neutral fraction of 0.25 rather than 0.45.  }
\label{HankHII_obs_SKA_0p75}
\end{center}
\end{figure*}

\begin{figure*}
\begin{center}
\includegraphics[width= 7.4cm]{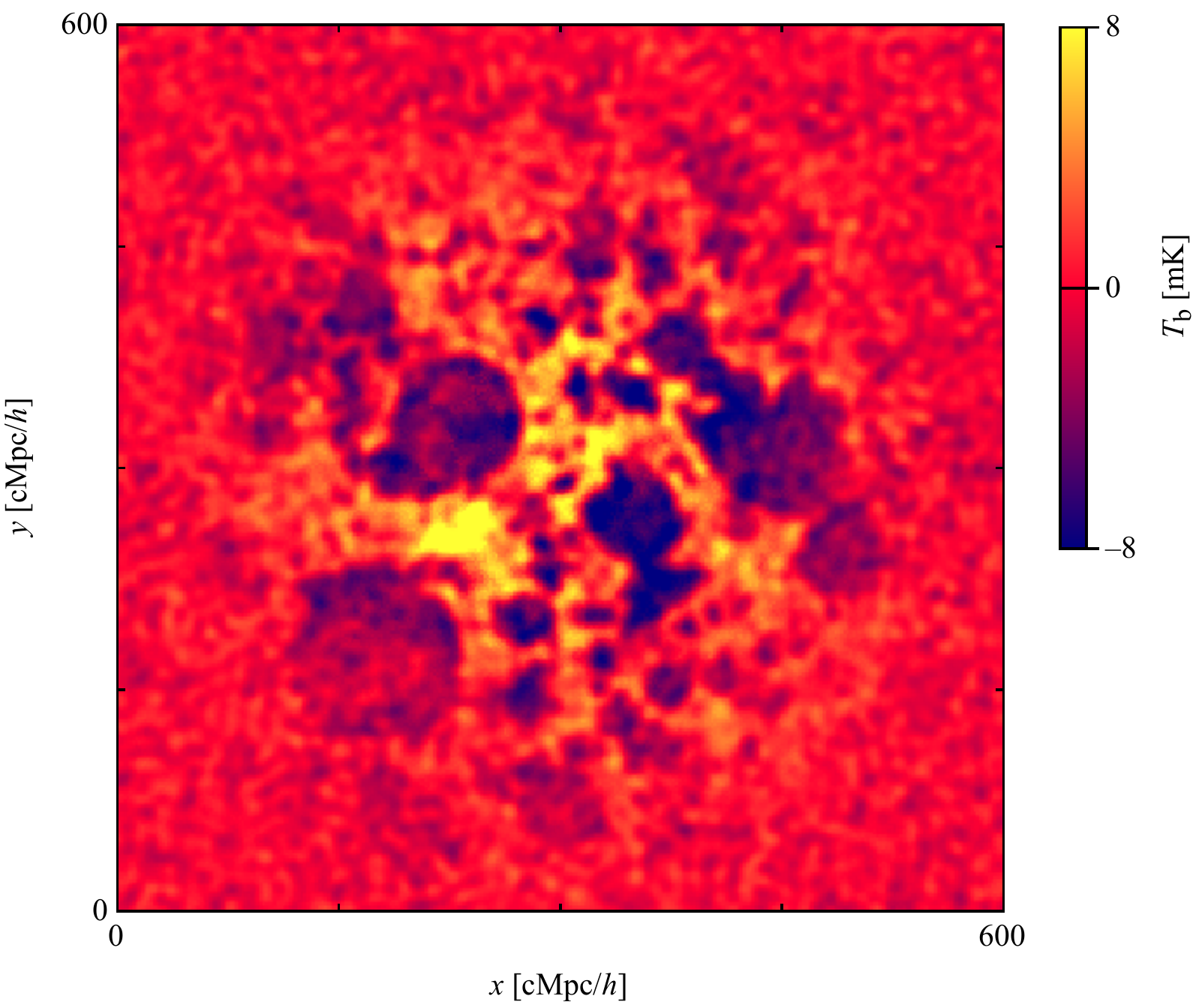}
\includegraphics[width= 7.4cm]{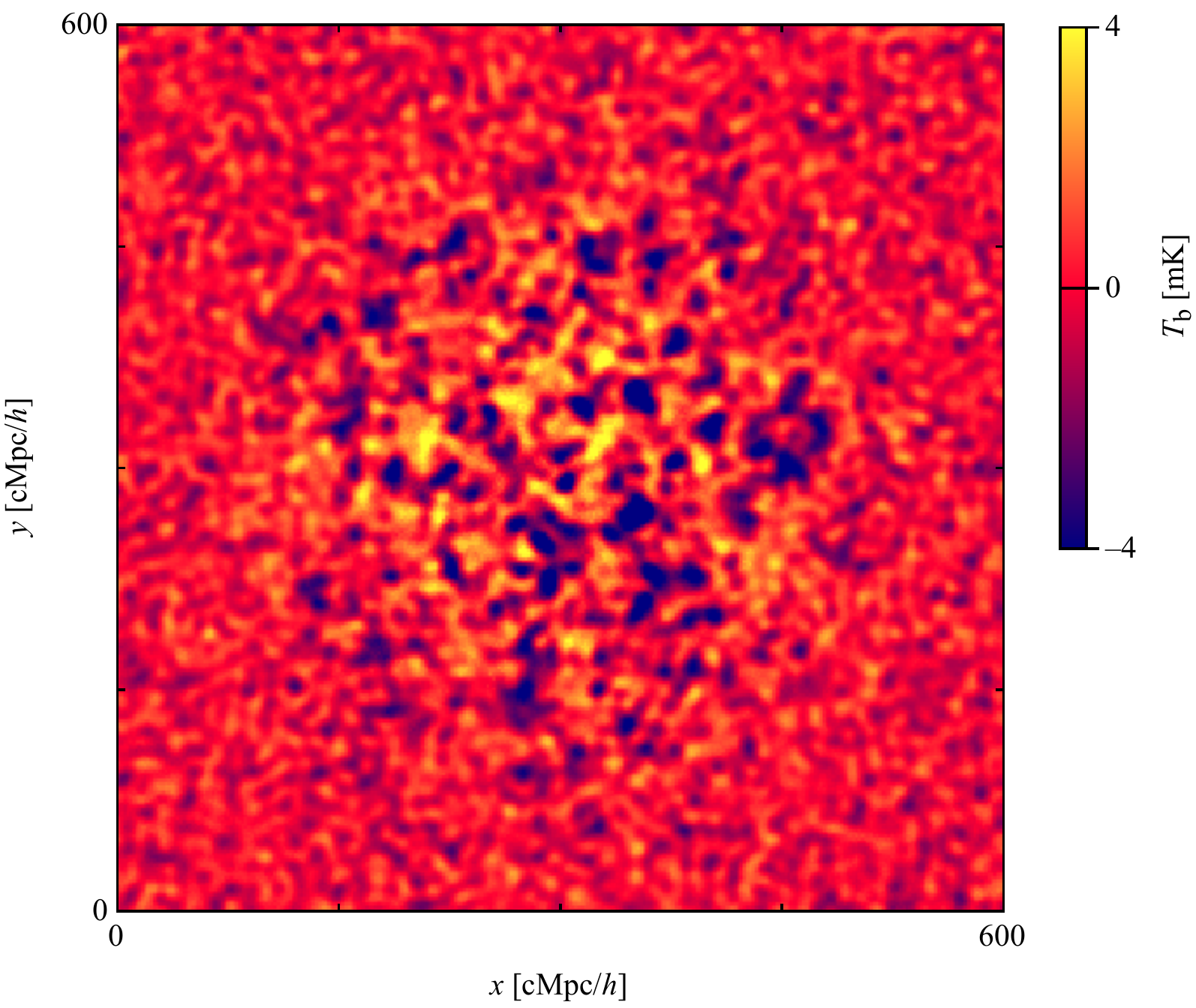}
\caption{\label{HankHII_obs_SKA}Simulations of the SKA response to the ionisation structure (without a primary beam correction) in the GiggleZ 1~Gpc$/h$ simulation in which star formation is assumed to  include an efficient SNe feedback (left) and to proceed in the absence of a strong SNe feedback (right). In both cases reionisation is assumed to have progressed to a neutral fraction of 0.45. For SKA, the sensitivity is assumed to increase by a factor of 4 (increasing the radius of the core by a factor of 2, and the density of stations in the arms by a factor of 4). The model observation has a depth of 7~Mpc$/h$, which corresponds to 0.6 MHz along the line-of-sight (at a central frequency of 172 MHz). }
\end{center}
\end{figure*}

Foreground emission, from Galactic synchrotron and extra-galactic point sources, will provide the largest challenge to measurement of the signal from reionisation.  In this discussion we assume that resolved sources can be successfully removed, which should be practical at the proposed sensitivity of SKA1-LOW~\citep{Liu2009}. We estimate the effects of removing the diffuse Galactic foregrounds, which can lower the contrast of observed images~\citep{GWPO08} by modelling the foreground continuum using a 2nd-order polynomial in the logarithm of frequency which is appropriate in the absence instrumental polarisation leakage~\citep{GWPO08,GGW2011}. We perform fits with this functional form along the line-of-sight within a 20 MHz bandwidth for each spatial pixel in the simulated image cube. We then subtract the best fit, leaving residual fluctuations around the foreground emission, which will include any residual continuum foregrounds, instrumental noise and the reionisation signal. The lower-left and lower-right panels of Figure~\ref{HankHII_obs} show simulated maps following diffuse foreground removal assuming 1000~hr integrations with the base-line SKA1-LOW design, and an early deployment of SKA1-LOW for which the sensitivity is decreased by a factor of 2 respectively. Here the slices are of depth 14 Mpc$/h$, which corresponds to 1.2 MHz along the line-of-sight, and is equivalent to the full-width at half power (FWHP) of the synthesised beam. The large HII regions produced by a model with SNe feedback can be imaged well by the SKA1-LOW baseline design, and marginally imaged with an early deployment of SKA1-LOW having half the sensitivity of the baseline design. The consequence of removing the foreground continuum, together with the fact that interferometers do not make zero-spacing measurements, results in a decreased contrast between ionised and non-ionised regions, and a loss of power from large-scale modes \citep{GWPO08}.

In Figure~\ref{HankHII_obs_NOSN}, we show examples of simulated maps for the case where HII regions are produced by galaxies in which star formation is assumed to proceed in the absence of a strong SNe feedback. Panels in this figure correspond to those in Figure~\ref{HankHII_obs}. We find that while the configuration of the SKA1-LOW baseline design can observe the largest of the smaller HII regions generated by a galaxy formation model without SNe feedback, these will in practice be difficult to observe owing to noise and the effects of foreground subtraction which lower the contrast of the observed HII regions. The SKA1-LOW baseline design appears to be the minimum configuration necessary to image HII regions generated by this second galaxy formation model during the early-to-mid portions of the reionisation era. Figure~\ref{HankHII_obs_SKA_0p75} shows the corresponding maps in the  scenario of galaxy formation without strong SNe feedback, where reionisation is more advanced with a smaller neutral fraction of 0.25. The larger HII regions generated later in reionisation could be resolved by SKA1-LOW. Note that although the HII regions in Figure~\ref{HankHII_obs_SKA_0p75} are of similar size to the case of galaxy formation with SNe feedback at an earlier phase of reionisation (Figure~\ref{HankHII_obs}), the sensitivity of SKA1-LOW would be sufficient to detect the excess of small-scale fluctuations seen in Figure~\ref{HankHII_obs_SKA_0p75}, and would therefore provide a discriminant between these models.

The left and right panels of Figure~\ref{HankHII_obs_SKA} show simulated maps following diffuse foreground removal assuming 1000~hr integrations with an SKA for which the sensitivity is increased by a factor of 4 over the SKA1-LOW baseline design. This sensitivity is achieved by increasing the density of stations in the spiral arms and size of the core (since increasing the density of the core would lead to a station filling factor greater than unity). As a result of the reduced FWHP of the synthesised beam, the model slice has a smaller depth of 7.4 Mpc$/h$, which corresponds to 0.45 MHz along the line-of-sight (at a central frequency of 172MHz). We find that the greater sensitivity and resolution of an SKA would allow more detailed imaging of HII regions generated by both the cases of galaxy formation with and without strong SNe feedback.

\section{Sharpness of HII regions}

Quasars have a harder spectrum than star forming galaxies, and this leads to thicker ionising fronts than is the case for a starburst driven HII region \citep{zaroubi2005,kramer2008}. With sufficient angular resolution the contribution of hard ionising sources, such as mini-quasars to reionisation could therefore be inferred from the structure of ionising fronts at the edge of HII regions \citep{tozzi2000}.  Similarly, the distribution of ionising sources surrounding a massive galaxy at the centre of an HII region prevents the boundaries of such HII regions from being sharp when viewed at finite resolution \citep{WyitheLoeb2007}. Rather, the clustering of sources near massive galaxies results in a spatially averaged neutral fraction that rises gradually towards large radii from an interior value near zero. As a result, a neutral hydrogen fraction corresponding to the global background value is typically reached only at a distance of 2--5 times the radius of the HII region around the central massive galaxy. This will lead to HII regions that look to have smooth edges unless observed at very high resolution. While detailed simulations remain to be done, inspection of Figures~\ref{HankHII_obs}--\ref{HankHII_obs_SKA} implies that this science will require the sensitivity and resolution of SKA.

\section{Field of view for SKA1-LOW}

Direct imaging of the ionisation structure on small scales during reionisation is challenging because of reduced sensitivity at high resolution.  However simulations suggest that very large structures of HI, with sizes of up to 100 Mpc will still be present during the later stages of reionisation~\citep{zaroubi2012}. Such large areas of patchy reionisation in the IGM result from the clustering of the large-scale structure on scales of up to $\sim120h^{-1}$ Mpc, or $\sim1$ degree (see Figure~\ref{HankHII}). Detection of these large-scale features may be possible at moderate significance with first generation arrays including LOFAR, and will be valuable for answering many cosmological questions~\citep{zaroubi2012}. However these expected large-scale features imply that the field of view for reionisation experiments performed with SKA1-LOW should have a size of at least several  degrees.

At these very large scales, the light travel time can become comparable to the Hubble time, implying that light-cone effects become important. Thus, while HII regions can become arbitrarily large in a simulation at fixed proper time (as may be seen in Figure~\ref{HankHII}) during the brief period of overlap at the end of reionisation, the combined constraints of cosmic variance and light travel time imply a maximum observed HII region size at the end of the overlap epoch (as may be seen in Figure~\ref{Wyithe1}). This maximum size is found to have a value of $\sim100$ Mpc~\citep{wl2004b}. In agreement with the simulations of  \citet{zaroubi2012}, this implies that reionisation experiments with SKA1-LOW should be sensitive to a characteristic angular scale of $\sim1$ degree for detection of the largest-scale 21~cm flux fluctuations near the end of reionisation, and have a field of view sufficiently large to image these features.

\section{Summary}

The properties of the ionisation structure of the IGM are sensitive to the unknown galaxy formation physics that prevailed during reionisation. This ionisation structure introduces non-Gaussian statistics into the redshifted 21~cm fluctuation amplitudes which can only be studied through tomographic imaging, and will clearly discriminate between different galaxy formation scenarios. Imaging the ionisation structure of the IGM during reionisation is therefore a key goal for the SKA. As an example, we have shown that the SKA1-LOW baseline design with a 1~km diameter core would resolve HII regions expected from galaxy formation models that include strong feedback on low-mass galaxy formation. However detailed imaging of the smaller HII regions that result from galaxy formation in the absence of SNe feedback may require the greater sensitivity of SKA, particularly in the early-to-mid phases of reionisation. In addition to having baselines long enough to resolve the typical HII regions, the field of view for SKA1-LOW reionisation experiments should be at least several degrees across in order to image the largest HI structures towards the end of reionisation. The baseline design with 35~m diameter stations has a field of view within a single primary pointing which is marginally sufficient for this purpose.

\pagebreak

\bibliographystyle{apj_long_etal}

\bibliography{SKA_Wyithe}

\end{document}